\documentclass[english,pra,twocolumn,groupedaddress,reprint]{revtex4}
\usepackage{graphicx}
\usepackage{amsfonts}
\usepackage[T1]{fontenc}
\usepackage[latin9]{inputenc}
\usepackage{color}
\usepackage{amsmath}
\usepackage{amssymb}
\usepackage{esint}
\usepackage{comment}
\usepackage{bbold}
\usepackage{tikz}
\usepackage{minitoc}
\usepackage{bm}

\begin{document}

\title{Pure optomechanical dark mode}
\author{Tao Wang\footnote{suiyueqiaoqiao@163.com} $^{1,2}$, Tie Wang $^{3}$, Chang-Bao Fu $^{2}$ and Xue-Mei Su\footnote{suxm@jlu.edu.cn} $^{1}$}
\affiliation{$^{1}$College of Physics, Jilin University, Changchun 130012, People's Republic of China}
\affiliation{$^{2}$College of Physics, Tonghua Normal University, Tonghua 134000, People's Republic of China}
\affiliation{$^{3}$Department of Physics, College of Science, Yanbian University, Yanji, Jilin 133002, People's Republic of China }

\begin{abstract}
Optomechanical dark mode plays a central role in effective mechanically-mediated conversion of two different optical fields. In this paper, two approaches are proposed to generate pure optomechanical dark mode, in which the optomechanical bright mode is absolutely eliminated even with small cooperativity and different losses of the two optical cavities. Only the optomechanical dark mode is left to facilitate quantum state transfer. This result provides a new perspective to utilize the dark state or dark mode for quantum information processing. ~~~~\newline
~~~~\newline
\end{abstract}

\maketitle

\section{Introduction}

In the field of quantum optomechanics, the motion of a mechanical oscillator can be coupled to one or many cavity fields via radiation pressure, thus the mechanical state can be prepared, probed and controlled by the cavity fields, or vice versa \cite{Review}. The cavity fields can be the optical field,  microwave field, radio field and other oscillation modes with different frequencies. This stimulates much interests in exploiting the use of the mechanical degree of freedom for applications in hybrid quantum network, quantum information processing and quantum communication. A lot of remarkable progresses have been made,  including cooling the mechanical resonator into the ground state \cite{Cooling1,Cooling2,Cooling3}, optomechanically induced transparency \cite{Transparency1,Transparency2}, optomechanical squeezing of light \cite{Lightsqueezing1,Lightsqueezing2,Lightsqueezing3}, strong entanglement between the cavity mode and the oscillator \cite{Entanglement}, squeezing of the mechanical state \cite{Motionsqueezing1,Motionsqueezing2,Motionsqueezing3}, conversion between different cavity modes \cite{Conversion1,Conversion2,Conversion3,Conversion4,Conversion5}, and preparation of a single-phonon state \cite{singlephonon1,singlephonon2}.

In almost all these progresses, thermal mechanical dissipation is one of the major obstacles, and preparing the mechanical oscillator into the ground state is the first critical step for utilizing the mechanical resonator \cite{Cooling1,Cooling2,Cooling3}. Only when it has been realized, many other techniques can be performed. However this direct approach is not the unique method for some quantum tasks, especially for quantum state conversion between two cavity fields with different frequencies. An indirect approach, exploiting the use of optomechanical dark mode, was proposed in Ref. \cite{tconversion1,tconversion2,tconversion3} and experimentally studied in Ref. \cite{Conversion1,Conversion2,Conversion3,Conversion4,Conversion5}. Two recent useful reviews on this topic can be found in Ref. \cite{Creview1,Creview2}.

Optomechanical dark mode is similar to the idea of dark state in coherent population
trapping and electromagnetically induced transparency in the $\Lambda$-style atoms, in which the dark state consists of the two low levels and can prohibit the excitation to the high level with large dissipation via destructive interference \cite{CPT,EIT}. In the existing scheme, two target cavity modes are simultaneously coupled to a common mechanical resonator, and can be hybridized to two normal modes: an optomechanical bright mode (OBM) and an optomechanical dark mode (ODM) . OBM is coupled to the mechanical resonator, so it suffers from the thermal fluctuations. ODM is decoupled from the thermal mechanical motion, so it can favor quantum state transfer with high fidelity.

Quantum state conversion includes the intra-cavity state transfer and the itinerant state transfer \cite{tconversion3}. In this paper, our purpose is to elaborate the idea of the pure ODM, so we only take the itinerant state transfer for example, in which the coupling strengths between the cavity modes and the mechanical oscillator are not varied. For itinerant state transfer, the key factor is that the excitation of the OBD can be greatly suppressed by optomechanically induced transparency \cite{tconversion1,tconversion3,Conversion2}.

Although the existing ODM scheme is very effective, many conditions should be satisfied. (1) The optomechanical cooperativity of the two cavity modes should be much larger than unity, so this scheme can not be used in the ultra-weak-coupling regime. (2) The ODM needs the same losses for the two cavity modes. (3) The OBM can not be eliminated absolutely, and the thermal mechanical noise can still destroy quantum state. In this paper we show that the OBM can be completely removed, and only the ODM is left, which is named pure ODM.

To obtain a pure ODM, another auxiliary cavity mode is needed, so the mechanical resonator is simultaneously coupled to three cavity modes. This three-terminal four-mode optomechanical setup recently attracts lots of attentions. The first proposal based on this setup is to generate strong entanglement between the two target cavity modes 1 and 2. Driving the auxiliary mode 3 at red-detuned sideband can greatly enhance the dissipation of the two target modes \cite{Fourmode1}. In addition, quantum-limited amplification with this setup was also discussed in Ref. \cite{Fourmode2}. Recently we found that prominent entanglement between the two target modes can be available under room-temperature \cite{Fourmode3}. In these results, one target mode is driven at blue-detuned  sideband, and another is driven at red-detuned sideband. In our paper, the two target modes are both driven at red-detuned sideband for state conversion.

We offer two methods to prepare the pure ODM. The first approach is to drive the auxiliary mode with two optical modes. One mode is strong at red-detuned sideband, and another is weak at the same frequency as the incident signal. The second approach is to use the auxiliary cavity mode to parametrically modulating the spring constant of the mechanical resonator at twice its natural frequency. This parametric modulation can induce the squeezing of the oscillator \cite{Parasqueezing1,Parasqueezing2}. Recently this interaction was exploited to amplify the single-photon nonlinearity \cite{Amplification1} and to realize phase-sensitive amplification and squeezing of an optical signal \cite{Amplification2}. In our paper, to emphasize the idea of the pure ODM, we discuss the itinerant state transfer at a classical level following the way in Ref. \cite{Conversion2}, which is suitable for converting a classical weak signal.

For the two approaches have the same target Hamiltonian, in section II we first introduce the target Hamiltonian and the ODM. In section III, we generate the pure ODM with weak driving light. In section IV, we introduce the pure ODM with parametric modulation. Although the two methods are different, the ultimate results are the same. The OBM is eliminated. The optomechanical cooperativity of the two cavity modes are equal for the ideal state conversion, which can reach unity, however they can be very small. The conclusion is given in section V.

\section{Target Hamiltonian and optomechanical dark mode}

The existing ODM scheme for itinerant state transfer has been clearly discussed in Ref. \cite{tconversion1,tconversion3,Conversion2}. In this section, we present a brief overview for the target Hamiltonian and the concept of ODM. This can be used to understand the differences between the existing ODM scheme and our pure ODM scheme. In addition, for consistency, the expressions of the OBM and the ODM are written in new forms. We discuss this problem at a classical level following the way in Ref. \cite{Conversion2} for simplicity, which is suitable for transferring a classical signal and also useful for understanding the nonclassical state conversion.

The optomechancal system consists of two target cavity modes 1, 2 and one mechanical resonator. The two cavity fields are simultaneously coupled to the same mechanical mode. The signal light enters into cavity 1 and comes out from cavity 2. The two target cavity modes are respectively driven by a strong control light at red-detuned sideband, and the coupling strengths between the cavity modes and the mechanical resonator can be greatly enhanced. Ultimately we have a linear-coupling interaction \cite{Creview1,Creview2}, and the target Hamiltonian for this system is
\begin{eqnarray}
H_{0}=H_{t}+H_{p},
\end{eqnarray}
\begin{eqnarray}
H_{t}&=&\hbar \omega_{m}(a^{\dag}_{1}a_{1}+a^{\dag}_{2}a_{2}+b^{\dag}b)   \nonumber\\
&&+\hbar G_{1}(a^{\dag}_{1}b+b^{\dag}a_{1})+\hbar G_{2}(a^{\dag}_{2}b+b^{\dag}a_{2}),
\end{eqnarray}
\begin{eqnarray}
H_{p}=i\hbar\sqrt{\kappa^{e}_{1}}( e^{-i\Delta t}a^{\dag}_{1}a_{1,in}- e^{i\Delta t}a^{\dag}_{1,in}a_{1}).
\end{eqnarray}
Here $H_{t}$ describes the energies of the two target cavity modes, the mechanical resonator and their interactions, and $H_{p}$ presents the driving interaction by the signal incident on the cavity mode 1. $a_{1}$, $a_{2}$ and $b$ are respectively the annihilation operators for the target cavity fields 1, 2 and the mechanical oscillator. $\omega_{m}$ is the mechanical frequency. If $\omega_{c,i}$ and $\omega_{l,i}$ (i=1, 2) are respectively the frequencies for the two cavity fields and the two driving lights on the two cavities, we have $\omega_{m}=\omega_{c,i}-\omega_{l,i}$ (i=1, 2). $G_{1}$ and $G_{2}$ are the effective optomechanical coupling rates between the two cavity modes 1, 2 and the mechanical resonator. $a_{1,in}$ is the annihilation operator for the incident signal light. If $\omega_{p}$ is the frequency of the signal light, $\Delta=\omega_{p}-\omega_{l,1}$ is the detuning between the signal light and the driving light 1. $\kappa^{e}_{1}$ is the effective output coupling rate of the cavity mode 1. It should be noticed that we work in the resolved-sideband regime $\kappa_{i}\ll \omega_{m}$ (i=1, 2) for any cavity mode, and the $\kappa_{i}$ (i=1, 2) are respectively the total decay rates of the cavity modes 1 and 2.

We define $\alpha_{1}=\langle a_{1}\rangle e^{-i\omega_{m}t}$, $\alpha_{2}=\langle a_{2}\rangle e^{-i\omega_{m}t}$, $\beta=\langle b \rangle e^{-i\omega_{m}t}$ and $\alpha_{1,in}=\langle a_{1,in}\rangle e^{-i\omega_{m}t}$, then the equations of motion for the target Hamiltonian can be described as follows
\begin{eqnarray}
\dot{\alpha}_{1}&=&-\frac{\kappa_{1}}{2} \alpha_{1} -iG_{1} \beta +\sqrt{\kappa^{e}_{1}} \alpha_{1,in}e^{-i \delta t}, \nonumber\\
\dot{\alpha}_{2}&=&-\frac{\kappa_{2}}{2} \alpha_{2} -iG_{2} \beta,   \nonumber\\
\dot{\beta}&=&-\frac{\gamma_{m}}{2}\beta-iG_{1}\alpha_{1} -iG_{2}\alpha_{2},
\end{eqnarray}
here $\gamma_{m}$ is the damping rate of the mechanical resonator, and $\delta=\Delta-\omega_{m}$. Let $\alpha_{1}=\alpha_{1-}e^{-i \delta t}$, $\alpha_{2}=\alpha_{2-}e^{-i \delta t}$ and $\beta=\beta_{-}e^{-i \delta t}$, then the steady state solution can be derived.

The OBM and the ODM can be defined as
\begin{eqnarray}
a_{B}=\frac{G_{1}a_{1}+G_{2}a_{2}}{G},  \nonumber\\
a_{D}=\frac{G_{2}a_{1}-G_{1}a_{2}}{G}.
\end{eqnarray}
Using these operators, $H_{t}$ can be rewritten as
\begin{eqnarray}
H_{t}&=&\hbar \omega_{m}(a^{\dag}_{B}a_{B}+a^{\dag}_{D}a_{D}+b^{\dag}b)   \nonumber\\
&&+\hbar G(a^{\dag}_{B}b+b^{\dag}a_{B}).
\end{eqnarray}
As shown in Eq. (6), the ODM is decoupled from the mechanical resonator, so it is immune to the thermal noise. The OBM is coupled to the mechanical oscillator with an effective coupling rate $G=\sqrt{G^{2}_{1}+G^{2}_{2}}$, so it suffers from the thermal noise. In the existing ODM scheme \cite{tconversion1,tconversion3,Conversion2}, the OBM can be greatly reduced due to optomechanically induced transparency.

According to Eq. (5), the amplitudes of the OBM and the ODM are $\alpha_{B}=\frac{G_{1}\alpha_{1_{-}}+G_{2}\alpha_{2_{-}}}{G}$ and $\alpha_{D}=\frac{G_{2}\alpha_{1_{-}}-G_{1}\alpha_{2_{-}}}{G}$. Thus we have
\begin{eqnarray}
\alpha^{(0)}_{B}&=&\frac{G_{1}}{G}\frac{D_{0}}{A+D_{0}} \frac{\sqrt{\kappa^{e}_{1}}}{\frac{\kappa_{1}}{2}-i\delta}\alpha_{1,in},  \nonumber\\
\alpha^{(0)}_{D}&=&(\frac{G_{2}}{G}-\frac{G_{1}}{G}\frac{B}{A+D_{0}})  \frac{\sqrt{\kappa^{e}_{1}}}{\frac{\kappa_{1}}{2}-i\delta}\alpha_{1,in},  \nonumber\\
\end{eqnarray}
where
\begin{eqnarray}
A&=&\frac{G^{2}_{1}}{\frac{\kappa_{1}}{2}-i\delta}+ \frac{G^{2}_{2}}{\frac{\kappa_{2}}{2}-i\delta}, \nonumber\\
B&=&\frac{G_{1}G_{2}}{\frac{\kappa_{1}}{2}-i\delta}- \frac{G_{1}G_{2}}{\frac{\kappa_{2}}{2}-i\delta},  \nonumber\\
D_{0}&=&\frac{\gamma_{m}}{2}-i\delta.
\end{eqnarray}
Here (0) denotes the existing ODM scheme.

For $\delta=0$, Eq. (7) can be reduced as
\begin{eqnarray}
\alpha^{(0)}_{B}(\delta=0)=\frac{G_{1}}{G} \frac{1}{1+C_{1}+C_{2}} \frac{2\sqrt{\eta_{1}}}{\sqrt{\kappa_{1}}} \alpha_{1,in},
\end{eqnarray}
\begin{eqnarray}
\alpha^{(0)}_{D}(\delta=0)=\frac{G_{2}}{G} \frac{1+\frac{\kappa_{1}}{\kappa_{2}}C_{1}+C_{2}}{1+C_{1}+C_{2}} \frac{2\sqrt{\eta_{1}}}{\sqrt{\kappa_{1}}} \alpha_{1,in}.
\end{eqnarray}
Here $C_{i}=\frac{4G^{2}_{i}}{\gamma_{m}\kappa_{i}}$ (i=1, 2) are the optomechanical cooperativity of the two cavity fields. We define $\eta_{i}=\frac{\kappa^{e}_{i}}{\kappa_{i}}$ for the output coupling ratios of the two cavity modes 1, 2.
Using the input-output relation, the outgoing signal is $\alpha_{2,out}=\sqrt{\kappa^{e}_{2}}\alpha_{2-}$. For $\alpha_{2-}=\frac{G_{2}\alpha_{B}-G_{1}\alpha_{D}}{G}$, the cavity mode-conversion efficiency at $\delta=0$ can be given by
\begin{eqnarray}
\chi^{(0)}=\frac{|\alpha_{2,out}|^{2}}{|\alpha_{1,in}|^{2}}=\eta_{1}\eta_{2}\frac{4C_{1}C_{2}}{(1+C_{1}+C_{2})^{2}}.
\end{eqnarray}
Eq. (9)$-$(11) are the main results in the existing ODM scheme \cite{tconversion1,tconversion2,tconversion3,Conversion1,Conversion2,Conversion3,Conversion4,Conversion5}. For overcoming the thermal noise, $\kappa_{1}=\kappa_{2}$ and $C_{1}=C_{2}\gg 1$ should be satisfied.

\section{Pure ODM with weak light driving}

Now we elaborate that the ODM can be eliminated completely via introducing a new control parameter. We introduce an auxiliary cavity mode 3, which is also coupled to the mechanical oscillator. The critical idea in this setup is to control the two target modes 1, 2 with the cavity mode 3 via the mechanical resonator. This three-terminal four-mode optomechanical setup has been recently discussed in Ref. \cite{Fourmode1,Fourmode2,Fourmode3} for strong entanglement and quantum-limited amplification. In this paper, we use the auxiliary mode to control quantum state transfer for the first time. The cavity mode 3 are driven by two control light. One is strong at red-detuned sideband for increasing the interaction between the cavity mode 3 and the mechanical resonator, and another is weak at the same frequency as the signal light. The cavity mode 3 is also at the resolved-sideband regime. 

The Hamiltonian can be described as
\begin{eqnarray}
H_{1}=H_{0}+H_{a1},
\end{eqnarray}
where
\begin{eqnarray}
H_{a1}&=&\hbar \omega_{m} a^{\dag}_{3}a_{3}+\hbar G_{3}(a^{\dag}_{3}b+b^{\dag}a_{3}) \nonumber\\
&+&i\hbar\sqrt{\kappa^{e}_{3}}( e^{-i\Delta t}a^{\dag}_{3}a_{3,in}- e^{i\Delta t}a^{\dag}_{3,in}a_{3}).
\end{eqnarray}
The auxiliary Hamiltonian $H_{a1}$ describes the energy of the cavity mode 3, the interaction between cavity mode 3 and the mechanical resonator, and the driving interaction by the weak control light. $a_{3}$ is the annihilation operator for the cavity field 3. If $\omega_{c,3}$ and $\omega_{l,3}$ are respectively the frequencies for the cavity field 3 and the strong driving light on it, $\omega_{m}=\omega_{c,3}-\omega_{l,3}$ should be also satisfied. $G_{3}$ is the effective optomechanical coupling rate between the cavity mode 3 and the mechanical resonator. $a_{3,in}$ is the annihilation operator for the weak control light. If $\omega_{w}$ is the frequency of the weak light, $\Delta=\omega_{w}-\omega_{l,3}$ is the detuning between the weak light and the driving light 3. $\kappa^{e}_{3}$ is the effective output coupling rate of the cavity mode 3.

Following the same process for the target Hamiltonian, we also define $\alpha_{3}=\langle a_{3}\rangle e^{-i\omega_{m}t}$ and  $\alpha_{3,in}=\langle a_{3,in}\rangle e^{-i\omega_{m}t}$, then the equations of motion for $H_{1}$ are
\begin{eqnarray}
\dot{\alpha}_{1}&=&-\frac{\kappa_{1}}{2} \alpha_{1} -iG_{1} \beta +\sqrt{\kappa^{e}_{1}} \alpha_{1,in}e^{-i \delta t}, \nonumber\\
\dot{\alpha}_{2}&=&-\frac{\kappa_{2}}{2} \alpha_{2} -iG_{2} \beta,   \nonumber\\
\dot{\beta}&=&-\frac{\gamma_{m}}{2}\beta-iG_{1}\alpha_{1} -iG_{2}\alpha_{2} -iG_{3}\alpha_{3},   \nonumber\\
\dot{\alpha}_{3}&=&-\frac{\kappa_{3}}{2} \alpha_{3} -iG_{3} \beta +\sqrt{\kappa^{e}_{3}} \alpha_{3,in}e^{-i \delta t}.
\end{eqnarray}
Here $\kappa_{3}$ is the total decay rate of the cavity mode 3.

Thus we can obtain the expressions of the OBM and the ODM for the weak driving case as follows
\begin{eqnarray}
\alpha^{(1)}_{B}&=&\frac{G_{1}}{G}\frac{D_{1}}{A+D_{1}}\frac{\sqrt{\kappa^{e}_{1}}}{\frac{\kappa_{1}}{2}-i\delta}\alpha_{1,in}  \nonumber\\
&&-\frac{G_{3}}{G}\frac{A}{A+D_{1}}\frac{\sqrt{\kappa^{e}_{3}}}{\frac{\kappa_{3}}{2}-i\delta}\alpha_{3,in}, \nonumber\\
\alpha^{(1)}_{D}&=&(\frac{G_{2}}{G}-\frac{G_{1}}{G}\frac{B}{A+D_{1}})\frac{\sqrt{\kappa^{e}_{1}}}{\frac{\kappa_{1}}{2}-i\delta}\alpha_{1,in}  \nonumber\\
&&-\frac{G_{3}}{G}\frac{B}{A+D_{1}}\frac{\sqrt{\kappa^{e}_{3}}}{\frac{\kappa_{3}}{2}-i\delta}\alpha_{3,in},
\end{eqnarray}
where
\begin{eqnarray}
D_{1}&=&(\gamma_{m}-i\delta)+\frac{G^{2}_{3}}{\frac{\kappa_{3}}{2}-i\delta}.
\end{eqnarray}
Here (1) denotes the pure ODM scheme with weak driving.

For $\delta=0$, Eq. (15) can be reduced as
\begin{eqnarray}
\alpha_{B}^{(1)}(\delta=0)&=&\frac{G_{1}}{G}\frac{1+G_{3}}{1+G_{1}+G_{2}+G_{3}}\frac{2\sqrt{\eta_{1}}}{\sqrt{\kappa_{1}}}\alpha_{1,in}  \nonumber\\
&&-\frac{G_{3}}{G}\frac{G_{1}+G_{2}}{1+G_{1}+G_{2}+G_{3}}\frac{2\sqrt{\eta_{3}}}{\sqrt{\kappa_{3}}}\alpha_{3,in},  \nonumber\\
~~~~~~
\end{eqnarray}
and
\begin{eqnarray}
\alpha_{D}^{(1)}(\delta=0)&=&\frac{G_{2}}{G}\frac{1+\frac{\kappa_{1}}{\kappa_{2}}C_{1}+C_{2}+C_{3}}{1+C_{1}+C_{2}+C_{3}}\frac{2\sqrt{\eta_{1}}}{\sqrt{\kappa_{1}}}\alpha_{1,in} \nonumber\\
&&-\frac{G_{2}G_{3}}{GG_{1}}\frac{C_{1}-\frac{\kappa_{1}}{\kappa_{2}}C_{1}}{1+C_{1}+C_{2}+C_{3}}\frac{2\sqrt{\eta_{3}}}{\sqrt{\kappa_{3}}}\alpha_{3,in}. \nonumber\\
~~~~~~~
\end{eqnarray}
Here $\eta_{3}=\frac{\kappa^{e}_{3}}{\kappa_{3}}$ is the output coupling ratio of the cavity mode 3. It is clear that, with assistance of the weak driving light, the OBM in Eq. (17) can be zero if the following condition is satisfied
\begin{eqnarray}
\frac{2\sqrt{\eta_{3}}}{\sqrt{\kappa_{3}}}\alpha_{3,in}= \frac{G_{1}}{G_{3}}\frac{1+C_{3}}{C_{1}+C_{2}}\frac{2\sqrt{\eta_{1}}}{\sqrt{\kappa_{1}}}\alpha_{1,in}.
\end{eqnarray}
Then the ODM can be further reduced as
\begin{eqnarray}
\alpha_{D}^{(1)}(\delta=0)&=&\frac{G_{2}}{G}\frac{\frac{\kappa_{1}}{\kappa_{2}}C_{1}+C_{2}}{C_{1}+C_{2}}\frac{2\sqrt{\eta_{1}}}{\sqrt{\kappa_{1}}}\alpha_{1,in},
\end{eqnarray}
so it is in fact independent of the thermal noise even the losses of the two target cavity modes are not equal. The cavity mode-conversion efficiency at $\delta=0$ can be given by
\begin{eqnarray}
\chi^{(1)}=\eta_{1}\eta_{2}\frac{4C_{1}C_{2}}{(C_{1}+C_{2})^{2}},
\end{eqnarray}
which is also immune to the mechanical dissipation, and when $C_{1}=C_{2}$ and $\eta_{1}\eta_{2}=1$, the conversion efficiency can approach unity. Here the $C_{1}$ and $C_{2}$ do not need to be much larger than unity, so  $G_{1}$ and $G_{2}$ can be very small.

\section{Pure ODM with parametric modulation}

We also notice that parametric modulation of the mechanical resonator can be also used to remove the OBM. Ref. \cite{Amplification2} pointed out that the unusual dynamics in optomechanics with mechanical parametric driving has interesting results for optomechanically induced transparency, and the cavity spectra function can be negative. Here we focus on the point that the cavity spectra function is zero.

The auxiliary mode 3 can be used to parametrically modulating the spring constant of the mechanical oscillator at twice the mechanical frequency \cite{Amplification1}, the effective Hamiltonian for this setup is
\begin{eqnarray}
H_{2}=H_{0}+H_{a2},
\end{eqnarray}
\begin{eqnarray}
H_{a2}=-\frac{\lambda}{2}(e^{-2i\omega_{m}t}(b^{\dag})^{2}+e^{2i\omega_{m}t}(b)^{2}).
\end{eqnarray}
$H_{a2}$ describes the parametric modulation interaction of the mechanical oscillator. $\lambda$ is the mechanical parametric driving strength. The equations of motion for $H_{2}$ are
\begin{eqnarray}
\dot{\alpha}_{1}&=&-\frac{\kappa_{1}}{2} \alpha_{1} -iG_{1} \beta +\sqrt{\kappa^{e}_{1}} \alpha_{1,in} e^{-i \delta t}, \nonumber\\
\dot{\alpha}_{2}&=&-\frac{\kappa_{2}}{2} \alpha_{2} -iG_{2} \beta,   \nonumber\\
\dot{\beta}&=&- \frac{\gamma_{m} }{2} \beta-iG_{1} \alpha_{1} -iG_{2} \alpha_{2}+\lambda\beta^{*}.
\end{eqnarray}
Thus the expressions for the OBM and the ODM can be derived as
\begin{eqnarray}
\alpha^{(2)}_{B}&=&\frac{G_{1}}{G}\frac{D_{2}}{A+D_{2}} \frac{\sqrt{\kappa^{e}_{1}}}{\frac{\kappa_{1}}{2}-i\delta}\alpha_{1,in},  \nonumber\\
\alpha^{(2)}_{D}&=&(\frac{G_{2}}{G}-\frac{G_{1}}{G}\frac{B}{A+D_{2}})  \frac{\sqrt{\kappa^{e}_{1}}}{\frac{\kappa_{1}}{2}-i\delta}\alpha_{1,in},
\end{eqnarray}
where
\begin{eqnarray}
D_{2}&=&(\frac{\gamma_{m}}{2}-i\delta)-\frac{|\lambda|^{2}}{(\frac{\gamma_{m}}{2}-i\delta)+\frac{G^{2}_{1}}{\frac{\kappa_{1}}{2}-i\delta}+\frac{G^{2}_{2}}{\frac{\kappa_{2}}{2}-i\delta}}. \nonumber\\
\end{eqnarray}
Here (2) denotes the pure ODM scheme with parametric modulation.

When $\delta=0$, Eq. (25) can be reduced as
\begin{eqnarray}
\alpha^{(2)}_{B}(\delta=0)=\frac{G_{1}}{G} \frac{1-t}{1-t+C_{1}+C_{2}} \frac{2\sqrt{\eta_{1}}}{\sqrt{\kappa_{1}}} \alpha_{1,in},
\end{eqnarray}
\begin{eqnarray}
\alpha^{(2)}_{D}(\delta=0)=\frac{G_{2}}{G} \frac{1-t+\frac{\kappa_{1}}{\kappa_{2}}C_{1}+C_{2}}{1-t+C_{1}+C_{2}} \frac{2\sqrt{\eta_{1}}}{\sqrt{\kappa_{1}}} \alpha_{1,in},
\end{eqnarray}
where
\begin{eqnarray}
t=\frac{4|\lambda|^{2}}{\gamma^{2}_{m}(1+C_{1}+C_{2})}.
\end{eqnarray}

It is clear that Eq. (27) can be set to zero if $1-t=0$ or $\lambda=\frac{\gamma_{m}}{2}\sqrt{1+C_{1}+C_{2}}$ is satisfied. We can see that the ODM is simplified as
\begin{eqnarray}
\alpha_{D}^{(2)}(\delta=0)=\frac{G_{2}}{G}\frac{\frac{\kappa_{1}}{\kappa_{2}}C_{1}+C_{2}}{C_{1}+C_{2}}\frac{2\sqrt{\eta_{1}}}{\sqrt{\kappa_{1}}}\alpha_{1,in},
\end{eqnarray}
and the cavity mode-conversion efficiency at $\delta=0$ can be given by
\begin{eqnarray}
\chi^{(2)}=\eta_{1}\eta_{2}\frac{4C_{1}C_{2}}{(C_{1}+C_{2})^{2}}.
\end{eqnarray}
These results for the ODM and the conversion efficiency with parametric modulation are the same as that with the weak light driving. For the OBM is completely removed, the ODM is absolutely decoupled from the mechanical resonator. Other approaches to remove the OBM will offer the same results.

\section{Conclusion}

In this paper, we point out that the OBM can be absolutely eliminated for state conversion between two cavity fields, and only the ODM is left for the task. Two approaches are put forward to elaborate this idea. Although the methods are different, the results obtained are the same. For the best conversion efficiency, it only needs that the two optomechanical cooperativity are equal, that $C_{1}=C_{2}$, except for the OBM removing condition. These results are important for designing new optomechanical interfaces for hybrid quantum networks, especially can be used for the weak light control when single-photon-level quantum state conversion is considered. Our discussion is at a classical level, and it is directly suitable for converting the weak classical signal. Our results reveals that, in the pure ODM scheme, the quantum state conversion can be a photon-number conservation process, so it is also useful for converting a quantum signal.

In fact, we offer a new perspective to utilize the dark state or dark mode, which plays an important role in many quantum tasks, such as coherent population trapping and electromagnetically induced transparency \cite{CPT,EIT}. Optomechanical systems can be coupled to many other quantum systems except for various cavity modes, such as Bose-Einstein condensate \cite{Condensate}, superconducting qubits \cite{Squbit1,Squbit2,Squbit3} and nitrogen-vacancy centers in diamond \cite{NV1,NV2}. Thus our method can be also used to investigate the qubit-state transfer between such quantum systems via the mechanical oscillator \cite{transfer1,transfer2}.

Many works based on this paper should be further clarified. Our results can be applied to convert a classical weak signal. For nonclassical quantum state conversion and single-photon-level quantum state transfer, a full-quantum treatment should be given \cite{quantumtreatment}. The added noise and transfer fidelity in diverse approaches to remove the OBM should be discussed. For the ODM is indirectly affected by the thermal mechanical noise via the OBM, it can be expected that the pure ODM is still immune to the mechanical dissipation in the full-quantum analysis.

\section{ACKNOWLEDGMENT}

 This research is supported by National Natural Science Foundation of China (Grant No.11174109) and (Grant No. 11404242).

\end{document}